\documentclass{optica-article}

\journal{opticajournal} 

\articletype{Research Article}

\usepackage{physics}
\usepackage{siunitx}
\DeclareUnicodeCharacter{2009}{\,}
\DeclareUnicodeCharacter{03BB}{\ensuremath{\lambda}}

\begin{document}

\title{Low-nonlinearity advantage of hollow-core fiber for wavelength multiplexing of quantum entanglement and classical DWDM signals}

\author{Hikaru Shimizu,\authormark{1,5} Arisa Ohno,\authormark{1} Mao Miyasugi,\authormark{1} Yumeno Seki,\authormark{1} Naoyuki Funaoka,\authormark{1} Yukari Takada,\authormark{2} Aruto Hosaka,\authormark{2} Shun Fujii,\authormark{3} Hiroyuki Tsuda\authormark{1} and Masahiro Takeoka\authormark{1,4,*}}

\address{\authormark{1}Department of Electronics and Electrical Engineering, Faculty of Science and Technology, Keio University, 
3-14-1 Hiyoshi, Kohoku-ku, Yokohama 223-8522, Japan\\
\authormark{2}Information Technology R\&D Center, Mitsubishi Electric Corporation, Kamakura 247-8501, Japan\\
\authormark{3}Department of Physics, Faculty of Science and Technology, Keio University, Kohoku-ku, Yokohama 223-8522, Japan\\
\authormark{4}National Institute of Information and Communications Technology~(NICT), Koganei, Tokyo 184-8795, Japan}

\email{\authormark{5}
shimihika2357@keio.jp}
\email{\authormark{*}
takeoka@elec.keio.ac.jp}


\begin{abstract*} 
Coexistence of classical and quantum light in optical fiber is a key requirement for scalable quantum networks. 
However, it is fundamentally limited by nonlinear crosstalk from intense classical signals. 
Hollow-core fiber~(HCF), with its intrinsically low optical nonlinearity, offers a promising solution.
Here, we experimentally demonstrate the coexistence of polarization-entangled photons with dense wavelength-division multiplexed~(DWDM) classical signals with a total launch power of 11.2 dBm, and perform a direct comparison between equal-length single-mode fiber~(SMF) and photonic-bandgap-based HCF deployed in a campus network. 
We observe that HCF suppresses background noise by approximately two orders of magnitude compared to SMF, while preserving high-fidelity entanglement distribution.
Our results provide clear experimental evidence of the advantage of HCF and establish its potential as a transmission medium.
\end{abstract*}

\section{Introduction}
Quantum networks enable the quantum interconnection of distant quantum devices. 
Given these specialized roles, quantum networks are not expected to replace conventional classical communication infrastructure, but rather to coexist with it. 
Experimental demonstrations of quantum entanglement distribution, a key component of quantum networks, have been extensively reported over the past two decades~\cite{Marcikic04, Honjo07, Inagaki13, Yin17, Valivarthi16, Wengerowsky20, Joshi20, Sun17, Yu20, Liu24, Lago-Rivera21, Hanni25, Stolk24, Craddock24, Shimizu26}.
Most of these demonstrations have focused on achieving higher fidelity and higher distribution rates, and are therefore performed in carefully controlled environments with minimal noise. 
However, as future quantum networks are expected to coexist with conventional optical communication infrastructure, experimental studies on quantum communication coexisting with classical signals have recently begun to emerge.
An important point in this context is that the simultaneous transmission of quantum and classical signals over a shared channel does not severely degrade the performance of either communication.
In the field of quantum key distribution~(QKD), many studies have explicitly considered operation alongside classical communication~\cite{Peters09, Qi10, Eriksson19, FerreiraDaSilva14, Patel12, Kumar15, Wu25}.
Recent works have extended coexistence studies beyond QKD to the distribution of quantum entanglement~\cite{Thomas24, Fan23}.
In many studies, high-precision filtering systems are employed to protect quantum channels, and the wavelength selection and power levels of classical signals are carefully optimized to suppress crosstalk arising from nonlinear optical effects in conventional silica-based single-mode fibers (SMFs).
However, imposing such constraints on classical signals will be a challenge for increasing the number of channels and scaling up the network.  

Hollow-core fiber~(HCF)~\cite{Fokoua23}, which has attracted considerable attention as a next-generation transmission medium, has a range of advantageous properties, including low latency~\cite{Zhu20, Chen24}, high power capability~\cite{Cooper23, Mulvad22}, low dispersion~\cite{Liu19, Zhang22}, and low transmission loss in anti-resonant type HCFs~\cite{Petrovich25}.
In particular, its intrinsically low optical nonlinearity~\cite{Liu19, Zhang22, Hong25} provides a promising pathway to mitigate crosstalk-induced degradation of quantum channels and to enable more relaxed conditions for classical signals.
Despite the difficulties in fabrication, a few studies have already employed HCF for quantum communication~\cite{Antesberger24, Clark25}.
In particular, Ref.~\cite{Clark25} has achieved coexistence with classical signals in the same band.

Reference~\cite{Clark25} represents an important first study demonstrating that HCF can protect quantum channels. 
The performance was evaluated by estimating possible secret-key rates of entanglement-based quantum key distribution.  
Furthermore, it demonstrated the distribution of Bell states among multiple users, showing a way to the realization of HCF-based quantum networks.   
On the other hand, its comparison with conventional SMF was limited to simulations, and an experimental and quantitative evaluation of the actual advantage of HCF remains unclear.

In this study, we demonstrate the coexistence of classical DWDM communications and quantum entanglement transmission over deployed HCF and SMF under the same conditions, and thus realize their experimental and quantitative comparison. 
A quantum channel, where polarization-based Bell states are transmitted, is multiplexed with classical DWDM channels in the C-band. 
The nearest classical channel is 250~GHz away from the quantum channel.
These signals are transmitted through equal-length HCF/SMF fiber links. 
We observe noise suppression at the quantum channel of more than two orders of magnitude for HCF compared to SMF. 
The transmitted entangled states are directly evaluated by quantum state tomography, and we demonstrate the degradation of entanglement in SMF by the classical DWDM, whereas the entanglement is preserved in HCF under the same conditions. 
The fidelity of the observed state to the original Bell state was $0.912\pm 0.003$ with HCF and $0.517\pm 0.004$ with SMF.

\section{Experiment}
\begin{figure}[b]
    \centering
    \includegraphics[width=1\linewidth]{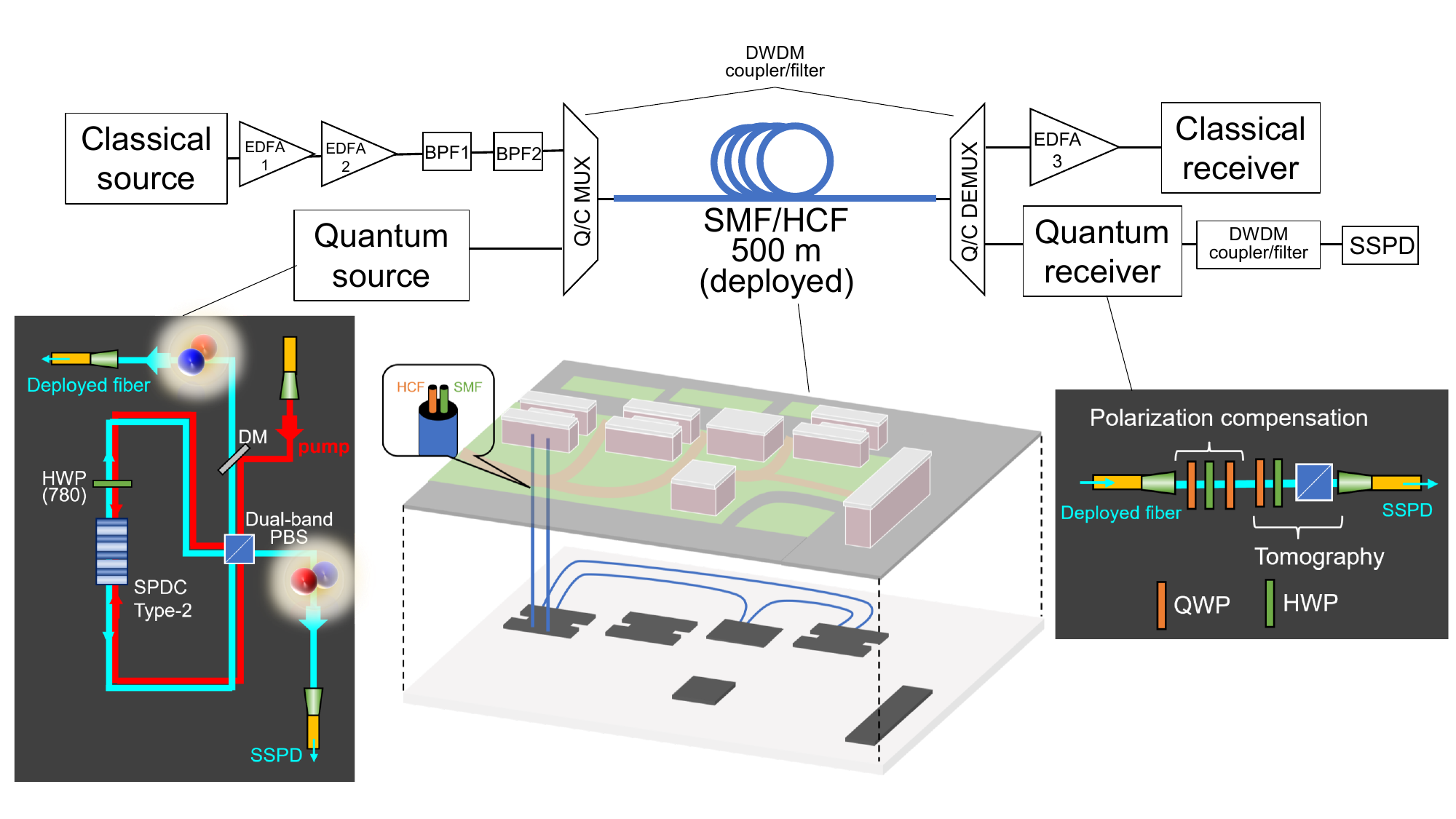}
    \caption{\raggedright Schematic of the whole setup.  The transmission fiber channel (both HCF and SMF) is deployed underground in the Shin-Kawasaki campus of Keio University, and the fiber length is about 500~m.
    EDFA: erbium-doped fiber amplifier, BPF: bandpass filter, SMF: single-mode fiber, HCF: hollow-core fiber, DM: dichroic mirror, PBS: polarizing beam splitter, HWP: half-wave plate, QWP: quarter-wave plate, SPDC: spontaneous parametric down-conversion, SSPD: superconducting single-photon detector.}
    \label{setup}
\end{figure}

Our experimental setup with the field fibers is illustrated in Fig.~\ref{setup}.
Quantum and classical communication signals are multiplexed and co-propagated through deployed fiber links.
To directly evaluate coexistence performance, the signals are transmitted through the same-length (approximately 500~m) SMF and HCF contained within the same cable installed at the Shin-Kawasaki campus of Keio University. 
Our HCF is based on a photonic-bandgap structure~\cite{Mukasa25}, which has various advantages of HCF, especially low nonlinearity, except for its relatively high optical loss. 
The fiber links are arranged in a loop configuration such that the transmitted signals return to the same room.
The measured loss is 0.6~dB for the SMF and 4.7~dB for the HCF.
The total input power of the classical signals is varied from 5.2 to 14.2 dBm.  
After transmission, each signal is separated by a demultiplexer and independently analyzed.

\begin{figure}[b]
    \centering
    \includegraphics[width=1\linewidth]{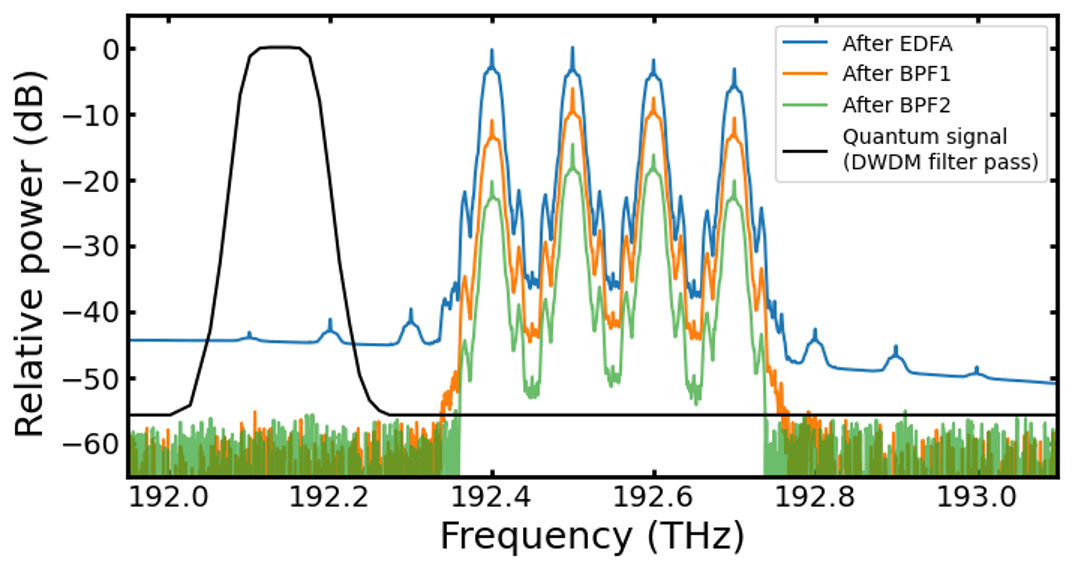}
    \caption{\raggedright Outputs of the optical spectrum analyzer and the transmission characteristic of our DWDM filter.}
    \label{wlassign}
\end{figure}

For the quantum signal, we used a polarization entanglement denoted as,
\begin{equation}
    \label{Bell}
    \ket{\Psi^+}=\frac{1}{\sqrt{2}}(\ket{HV}+\ket{VH}).
\end{equation}
The entanglement is generated by a Sagnac interferometer configuration that includes spontaneous parametric down conversion~(SPDC).
We use a type-2 periodically poled lithium niobate~(PPLN) crystal.
The pump light for SPDC is generated by second-harmonic generation~(SHG) of telecom-wavelength pulsed laser light.
Inside the interferometer, we used a dual-band polarizing beam splitter~(DPBS) to properly route both pump light and photons.
A half-wave plate~(HWP) placed before the crystal is designed to operate only at 780~nm while leaving the polarization of telecom-band light unchanged.
One qubit of the generated entanglement is immediately measured using a polarization quantum state tomography~(QST) setup consisting of a quarter-wave plate~(QWP), a HWP, and a PBS, followed by a superconducting nanowire single photon detector~(SSPD).
Before the SSPD, we put a DWDM filter to get a spectrally single-mode photon.
The other qubit is transmitted to the deployed fiber and multiplexed with classical signals.
After propagating through the fiber channel, the demultiplexed quantum signal is polarization-compensated with QWP-HWP-QWP waveplate sets and measured using the same QST setup.
Though the transmitted photon is already spectrally filtered by the demultiplexer, we placed an additional DWDM filter before the SSPD to suppress background noise from leaked classical signals and clearly evaluate noise from fiber nonlinear effects.
The photons generated by SPDC have a broader spectrum than the classical communication signals, so the filters determine their effective spectrum. 
In this experiment, we selected a spectral mode centered at approximately 192.15~THz (black curve in Fig.~\ref{wlassign}).

For classical communication, we used four DWDM channels whose central frequencies are 192.4, 192.5, 192.6, and 192.7~THz.
We modulated these classical signals using quadrature phase shift keying~(QPSK) at a symbol rate of 25~Gbaud and successfully demodulated them in the presence of the quantum signal.

The quantum and classical signals are multiplexed and demultiplexed using DWDM couplers and filters (identical components, designated by the direction of signal propagation).
The optical spectrum at various points in the system is monitored using an optical spectrum analyzer, as shown in Fig.~\ref{wlassign}. 
Sidebands generated by four-wave mixing (FWM) are clearly observed after the EDFA, confirming the presence of nonlinear optical processes. 
The increase towards lower frequencies is likely due to Raman scattering.
Since the spectrum of photons generated by SPDC is broad, the transmission spectrum of this filter directly determines the quantum signal spectrum. 
Assuming dense frequency allocation, the classical signal was placed as close as possible to the quantum signal, down to the lowest frequency at which the SSPD can still operate, even with SMF transmission.
To ensure that the comparison reflects only noise generated within the deployed fiber, we insert two bandpass filters (BPFs) before multiplexing. 
This filtering allows to independently evaulate the impact of fiber-induced nonlinear processes on the quantum channel.


\section{Results}
\begin{figure}[b]
    \centering
    \includegraphics[width=1\linewidth]{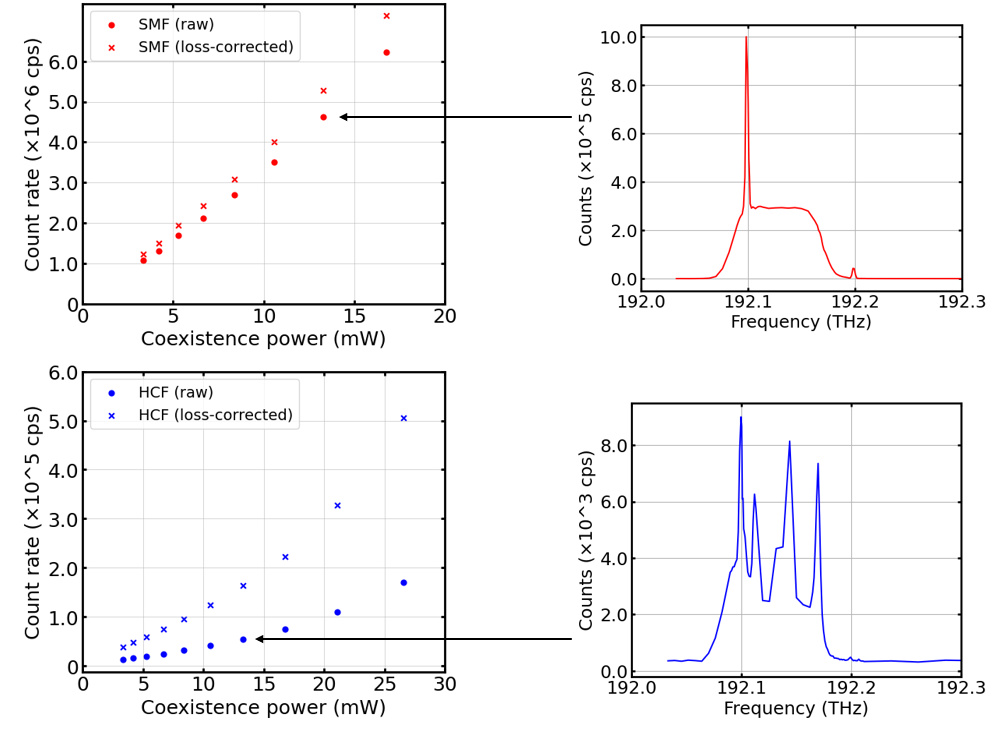}
    \caption{\raggedright Measured background noise for SMF (top) and HCF (bottom). The left panels show the SSPD count rates measured using the quantum-signal detection setup. The right panels show measurements obtained by replacing the DWDM filter with a wavelength-tunable filter.
}
    \label{BGnoise}
\end{figure}

First, we measure the background noise at the quantum channel by varying the total input power of classical signals in Fig.~\ref{BGnoise}.
In this measurement, we did not input a quantum signal.
A clear difference is observed between the HCF and SMF. 
The measured count rates in HCF are significantly lower than those in SMF over the entire power range. 
We also measured the background noise spectrum by replacing the DWDM filter in front of the SSPD with a tunable bandpass filter for the input of 11.2 dBm (Fig.~\ref{BGnoise} right).
The noise levels measured in the HCF are approximately two orders of magnitude lower than those in the SMF. 
In the SMF spectrum, distinct peaks attributed to FWM are observed above a broadband background suggestive of Raman scattering. 
A few peaks and broadband background noise are also observed in the HCF spectrum, but both are substantially weaker.
This result directly demonstrates the suppression of fiber-induced nonlinear noise in HCF.
One possible concern is that the observed background counts may originate from leakage of classical communication signals rather than fiber-induced nonlinear optical effects, and that the apparent advantage of HCF could simply result from differences in transmission loss between the two fiber types. 
To address this concern, we also plotted the measured background count rates, corrected for channel transmittance~(loss-corrected). 
The substantial difference remains even after loss correction, indicating that the observed noise cannot be attributed to leakage from classical communication signals and that the observed noise reduction originates from intrinsic suppression of nonlinearity in HCF rather than from differences in transmission loss.

\begin{figure}[t]
    \centering
    \includegraphics[width=1\linewidth]{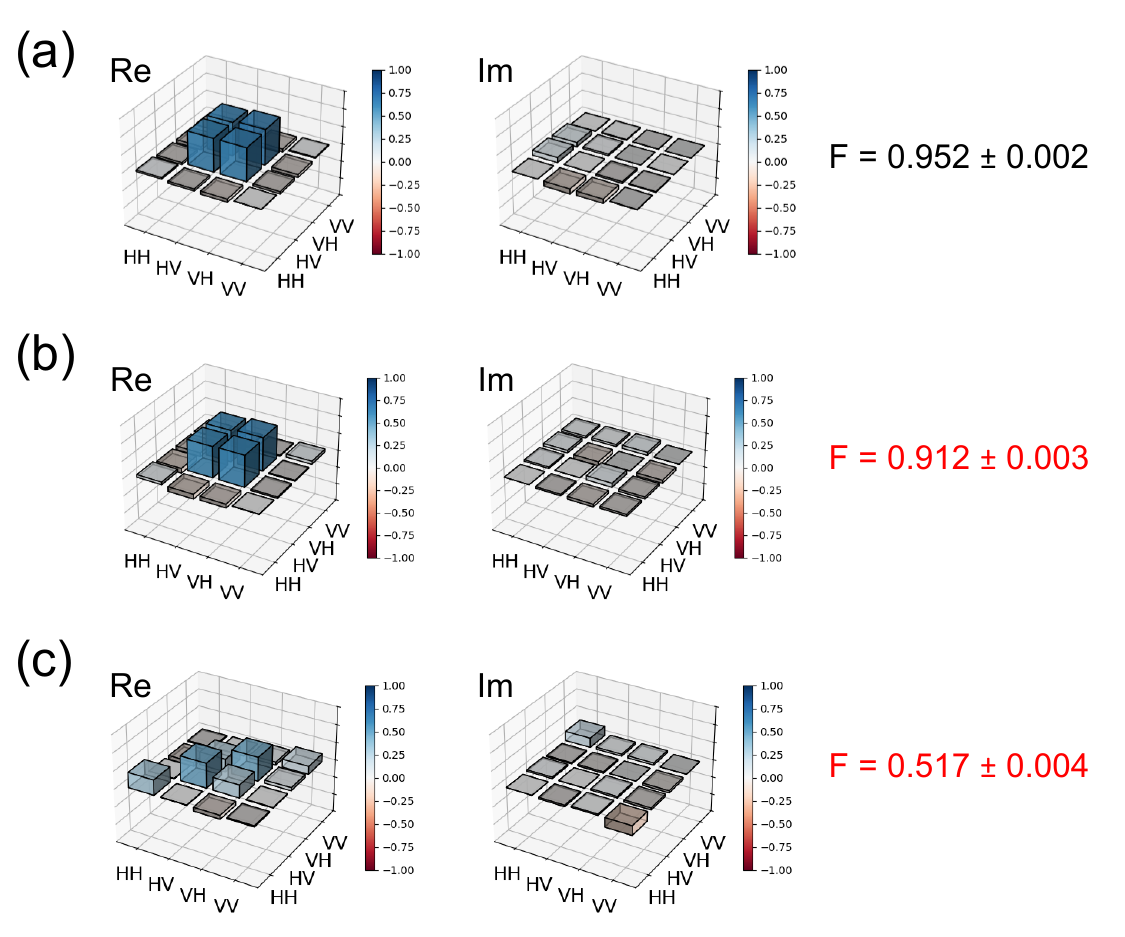}
    \caption{\raggedright Reconstructed density matrices of the polarization-entangled state (a) before transmission and after transmission together with classical communication signals through (b) HCF and (c) SMF. 
    The real and imaginary parts are shown in the left and right columns, respectively. 
    Fidelities to the target Bell state are calculated as F = $\langle \Psi^{+}|\rho|\Psi^{+}\rangle$, where $\rho$ is the reconstructed density matrix.}
    \label{result_tomography}
\end{figure}

We next evaluate the impact of fiber-induced nonlinear noise on entanglement preservation by quantum state tomography. 
In this experiment, we multiplex polarization-entangled states at 192.15~THz with classical DWDM signals with a total power of 11.2 dBm. 
Figure~\ref{result_tomography} shows the reconstructed density matrices of the locally generated polarization entanglement~(Fig.~\ref{result_tomography}(a)) and transmitted quantum states for both HCF~(Fig.~\ref{result_tomography}(b)) and SMF~(Fig.~\ref{result_tomography}(c)) links with classical coexistence.
In the absence of classical signals, high-fidelity entanglement is observed for both fiber types, with fidelities of $0.937 \pm 0.006$ and $0.939 \pm 0.002$ for HCF and SMF, respectively (not shown in Fig.~\ref{result_tomography}). 
This confirms that fiber transmission alone does not significantly degrade the entangled state.
However, under coexistence conditions with classical signals, a pronounced difference emerges. 
In the SMF transmission, the fidelity decreases significantly to $0.517 \pm 0.004$, indicating substantial entanglement degradation due to fiber-induced nonlinear noise. 
In contrast, the HCF maintains high fidelity of $0.912 \pm 0.003$, demonstrating robust entanglement preservation even under strong classical coexistence conditions.
These results directly confirm that suppressing nonlinear effects in HCF enables reliable transmission of polarization-entangled photons alongside high-power classical communication signals.

\section{Discussion}
Although a small residual background is still observed in the HCF link, this contribution is likely dominated by external factors, such as weak leakage of classical signals and nonlinear optical effects generated in short segments of conventional SMF that precede the HCF. 
Given the intrinsically low optical nonlinearity of HCF, nonlinear optical effects occurring within the HCF itself are expected to be negligible over the fiber length used in this experiment.

The observed power dependence suggests that, at the classical launch power used in this experiment, the in-band noise is likely dominated by broadband processes, including spontaneous Raman scattering, rather than by the discrete FWM peaks alone.
In silica fibers, the Stokes Raman gain typically reaches its maximum at a frequency downshift of approximately 13~THz from the pump frequency and extends over a broad spectral range~\cite{Agrawal11}. 
Therefore, in the present C-band wavelength configuration, the quantum channel is located on the rising tail of the Stokes Raman spectrum rather than near its maximum.
The wavelength configuration used in this experiment was chosen to emulate dense wavelength allocation while avoiding direct leakage of classical signals due to finite filter isolation. 
It also avoids the spectral region where the Raman contribution is expected to be strongest. 
Nevertheless, under this relatively favorable configuration for SMF coexistence, the SMF exhibited a large in-band noise background and failed to preserve entanglement.
These observations imply that robust coexistence in SMF would require more restrictive wavelength allocation, such as placing the quantum channel on the higher-frequency side of the classical channels, using a much larger spectral separation on the lower-frequency side, or reducing the launched classical power. 
In practical DWDM networks, where many classical channels may occupy the same transmission band, these constraints can significantly reduce the available spectral windows for quantum channels. 
In contrast, HCF suppresses the generation of classical-light-induced noise itself, thereby relaxing these wavelength-allocation constraints. 
This represents a key advantage of HCF for multiplexing quantum channels with dense classical optical communication channels. 

Finally, we discuss the meaning of the input power of classical DWDM in our experiment. 
We observe the advantage of HCF in the range from 3 to 30 mW of the total input power for the 4-ch classical DWDM. 
This means each channel's signal power is a few mW, which may be too high for practical communication. 
However, these total input powers suggest an important implication for future dense multiplexing of quantum signals and high-capacity optical communication. 
For example, a 100-ch DWDM with a total power of 13.2~mW (=11.2~dBm, the power used for our entanglement transmission experiment) means 132~$\mu$W/channel, which is a reasonable power for practical optical communication. 
Thus, our results strongly suggest a potential capability of HCF for the dense multiplexing of entanglement transmission and high-capacity optical channels.

\section{Conclusion}
In this work, we experimentally investigated the coexistence of polarization-entangled photons and classical DWDM signals over deployed SMF and HCF links. 
By directly comparing equal-length fiber links under identical launch conditions with a total classical input power of 11.2 dBm, we demonstrated that HCF significantly suppresses noise generated by nonlinear optical effects during transmission. 
The observed reduction in background counts persists even after correcting for transmission loss, confirming that the improvement originates from intrinsically reduced nonlinear optical effects rather than attenuation or leakage of classical signals.
Quantum state tomography further revealed that entanglement is severely degraded after transmission through SMF under coexistence conditions, while high-fidelity entanglement is preserved in HCF. 
These results provide clear experimental evidence that HCF enables robust transmission of entangled photons alongside high-power classical communication signals.
Our findings highlight HCF as a promising transmission medium for scalable quantum networks that coexist with conventional optical communication systems.

\begin{backmatter}
\bmsection{Funding}
Program for the Advancement of Next Generation Research Projects, Keio University; JST CRONOS (JPMJCS24N6); JST ASPIRE (JPMJAP2427); JST Moonshot R\&D (JPMJMS2061, JPMJMS226C); NEDO JPNP25014; MEXT Quantum Leap Flagship Program (Q-LEAP)(JPMXS0118067246); MIC (JPMI00316).

\bmsection{Acknowledgment}
This research was conducted at the Keio Future Optical Network Open Lab.
M.T. acknowledges the members of the Quantum Internet Task Force for the comprehensive and interdisciplinary discussions on the quantum internet.

\bmsection{Disclosures}
The authors declare no conflicts of interest.

\bmsection{Data Availability Statement}
 Data underlying the results presented in this paper are not publicly available at this time but may be obtained from the authors upon reasonable request.
\end{backmatter}

\bibliography{ref.bib}

\end{document}